\documentclass[10pt, conference, letterpaper]{IEEEtran}
\usepackage[english]{babel}
\usepackage{blindtext}
\usepackage[utf8]{inputenc}
\usepackage[dvipsnames]{xcolor}
\usepackage{verbatim}
\usepackage{hyphenat}
\usepackage{epstopdf}
\usepackage{graphics}
\usepackage{graphicx}
\usepackage{amsmath}
\usepackage{multirow}
\usepackage{tabularx}
\usepackage{float}
\usepackage{color}
\usepackage[linesnumbered,ruled]{algorithm2e}
\usepackage{amssymb}
\usepackage{array}
\usepackage{pbox}
\usepackage{makecell}
\usepackage{cite}
\usepackage{color,soul}

\newcolumntype{C}[1]{>{\centering\arraybackslash}m{#1}}
\newcommand{\nameOfSystem}{IsoRAN }
\newcommand{\nameOfSystemNoSpace}{IsoRAN}

\newcolumntype{Y}{>{\centering\arraybackslash}X}

\begin{document}

\title{IsoRAN: Isolation and Scaling for 5G RAN \\via User-Level Data Plane Virtualization}
\author{Nishant Budhdev, Mun Choon Chan, Tulika Mitra \\
        School of Computing, National University of Singapore \\ 
        \texttt{\{nishant,chanmc,tulika\}@comp.nus.edu.sg}
        }


\maketitle
\begin{abstract}

5G presents a unique set of challenges for cellular network architecture. The architecture needs to be versatile in order to handle a variety of use cases. While network slicing has been proposed as a way to provide such versatility, it is also important to ensure that slices do not adversely interfere with each other. In other words, isolation among network slices is needed. Additionally, the large number of use cases also implies a large number of users, making it imperative that 5G architectures scale efficiently.

In this paper we propose IsoRAN, which provides isolation and scaling along with the flexibility needed for 5G architecture. 
In IsoRAN, users are processed by daemon threads in the Cloud Radio Access Network (CRAN) architecture. Our design allows users from different use cases to be executed, in a distributed manner, on the most efficient  hardware to ensure that the Service Level Agreements (SLAs) are met while minimising power consumption. Our experiments show that IsoRAN handles users with different SLA while providing isolation to reduce interference. This increased isolation reduces the drop rate for different users from 42\% to nearly 0\% in some cases. Finally, we run large scale simulations on real traces to show the benefits for power consumption and cost reduction scale while increasing the number of base stations.

\end{abstract}

\section{Introduction}
\label{sec:introduction}





As we move towards 5G, many of the assumptions that have guided the previous generations of RAN design are no longer valid.
While previous generations of RAN look at supporting either voice or data traffic, 5G is expected to support a wide variety of users with very different requirements. 
Some of the principle use cases are enhanced mobile broadband\,(eMBB), ultra-reliable and low latency communication\,(uRLLC), and massive machine type communication\,(mMTC)~\cite{itu2018report}. These use cases demonstrate the diversity of 5G services, from eMBB services that aim to support Gigabit per second download speeds, to uRLLC services requiring high reliability and sub-millisecond signalling latency, to mMTC services targeting millions of devices within a kilometer square~\cite{3gpp38913}. 
To support these new demands, 5G RAN architecture needs to be versatile to accommodate different use cases and their varying requirements. Furthermore, the RAN also needs to provide isolation 
between these use cases.
Finally, the RAN should scale efficiently to support a large number of users within each use case\cite{ciscoVni}. To achieve these qualities, 5G RAN architecture proposals use a combination of Software Defined Networking\,(SDN) and Network Function Virtualization\,(NFV). 
SDN enables versatility as the control plane can support multiple use cases without changing the underlying data plane, while
NFV facilitate scaling and support custom data plane processing for different use cases by implementing hardware based services as software applications enabling the usage of general-purpose processors.

Virtualization also enables movement of network services to centralized clouds thereby providing isolation and scaling. Multiple works propose using such a centralized solution called CRAN\cite{alliance2013suggestions,ericsson2015cran,chang2017flexcran}. 
Such proposals involve splitting RAN functionality between distributed base stations and the centralized RAN, to enable multiplexing of resources at the centralized cloud to reduce total costs.
In cellular networks, CRAN can be further logically separated into different parts called slices.


\begin{figure}[!t]
\centering
    \includegraphics[scale=0.17]{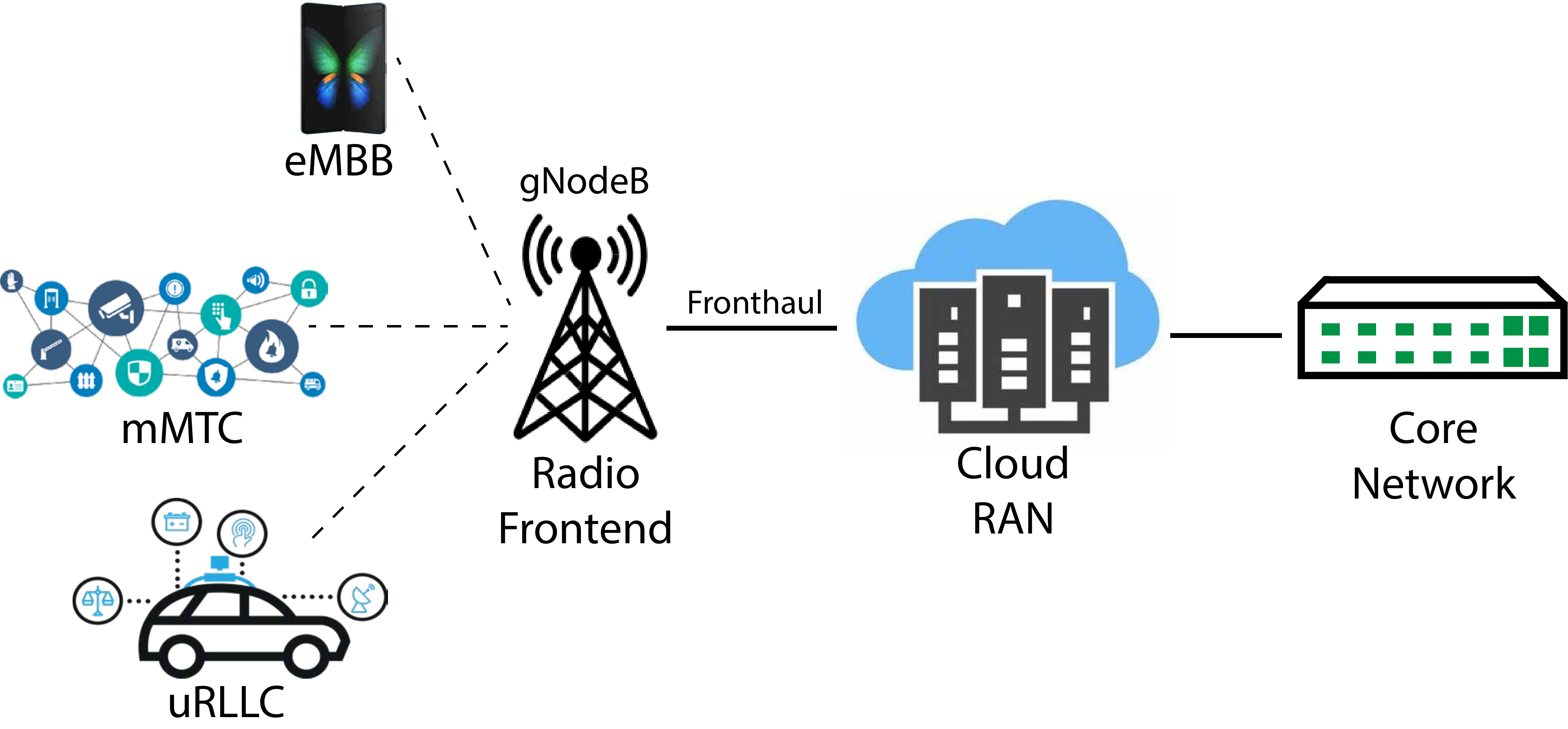} 
    \caption{5G CRAN architecture, supporting multiple use cases and RATs with different SLAs. }
    \vspace{-0.5cm}
    \label{fig:5gCranArchitecture}
\end{figure}

The use of slicing in RAN architecture has been proposed to provide versatility~\cite{an2017end,ngmnwhitepaper,foukas2017orion}.
However, while slicing improves versatility, it does not necessary provide sufficient protection from different services and users. The 3GPP \cite{3gpp38801} standard has stated that isolation between slices is critical for avoiding adverse impact due to other slices.
The most common slicing approach places each slice of a base station in a virtually isolated environment on a single host. 
As services such as eMBB (high throughput) and uRLLC (low latency) have very diverse requirements, placing them on the same host requires over-provisioning in order to minimize the adverse effects of interference due to sharing processing, memory and networking resources. As a result, slicing either may not scale well due to inefficient resource usage or unable to meet the service requirements if the issue of isolation is not addressed properly. 

To address the above mentioned challenges, we present \emph{IsoRAN}, a design for 5G RAN that can run on heterogeneous cloud architecture to provide the flexibility,  isolation, and scalability required to meet diverse throughout and latency requirements with lower cost.
The key idea of IsoRAN is the use of user-level data plane virtualization. 
This enables distributed processing in the RAN, similar to traditional cloud applications where processing is spread across large number of threads on multiple hosts.
In \nameOfSystemNoSpace, users attached to the same base station are not restricted to a single host and can now instead be placed anywhere in the cloud.
Furthermore, users with the same SLA when attached to different base stations, can be grouped together under a large slice, so as to minimize the effect of adverse interference and reduce network orchestration complexity.
On the other hand, users with different SLA can be processed on host configurations optimal for their processing.
These design principles are made practical due to two key advancements: (1) support for multipoint-to-multipoint routing in the fronthaul~\cite{cpri2019ecpri}, and (2) presence of general-purpose processors on smartNICs~\cite{netronomeNIC,mellanoxNIC,caviumNIC}.
Our contributions are as follow:
\begin{itemize}
    \item IsoRAN separates user-specific function such as coding/decoding, modulation/demodulation, interleaving etc., and cell-specific functionality such as synchronization, paging requests etc.
    \item IsoRAN is designed so that each user thread works on an independent data stream with no need for locking/synchronization among threads. This is done with minimum state sharing and shared states being read-only data in nature. 
    \item We have built a working prototype implementation of IsoRAN using OpenAirInterface (OAI)\footnote{https://www.openairinterface.org/} with user-level threading and parallelization in the data plane. 
\end{itemize}

Our evaluations on different hardware platforms show that the flexibility enabled by IsoRAN allows the host allocation framework to meet the SLA for different classes of services with lower power consumption. Finally, a large scale simulation using traffic traces collected over a 5 hour period spread over 15 days shows that IsoRAN can reduce both total power consumption and cost while meeting the SLAs.



The rest of the paper is organized as follow. Section \ref{sec:relatedwork} discusses related research proposals that address different requirements and we motivate our design in Section \ref{sec:Motivation}. 
We present the design of IsoRAN in in Section \ref{sec:systemdesign}. 
Section \ref{sec:evaluation} evaluates IsoRAN's flexibility, isolation and scaling. Finally we conclude in Section \ref{sec:conclusion}.

\section{Related Work}
\label{sec:relatedwork}

Early proposals of SDN and NFV were often limited to the core network owing to its similarity with wired networks. These works~\cite{jin2013softcell,banerjee2015scaling,costanzo2014openb,moradi2014softmow,pentikousis2013mobileflow} focus on implementation of dynamic control plane applications which address issues ranging from managing mobility to handling heterogeneous users. However, none of these works focus on the RAN thereby limiting the overall flexibility of these architectures. This can be attributed to the unique control plane and data plane dependencies in the RAN, which are often associated with stringent timing constraints.

Recently, a number of works have addressed this by separating the control plane and the data plane. The focus of this body of work~\cite{foukas2017orion}\cite{gudipati2013softran,Akyildiz2015,arslan2015software,chen2014softmobile,flexran} is on improving both flexibility and isolation in the RAN. These works build upon the concept of logically separate network called slices, with each slice capable of supporting a unique set of SLA. Another common principle among these works is the centralization of RAN processing. These works differ in the level of logical separation provided by slicing and the degree of centralization in RAN.

SoftRAN~\cite{gudipati2013softran} introduces a big-base station abstraction implemented by a software defined centralized control plane. The control plane functions are statically implemented either centrally or distributedly, based on their timing constraints and type of function performed. Contrary to this, FlexRAN~\cite{flexran} introduces an adaptive and flexible functional split for the RAN. Additionally, it is one of the earliest works which involve an end-to-end implementation of the RAN architecture with a centralized control plane. However, FlexRAN does not deal with heterogeneity of users from the perspective of wireless resources or their processing. Recently, Orion\cite{foukas2017orion} proposed a RAN slicing system which enables flexible customization of slices to accommodate varying needs of different use cases. The architecture implements a hypervisor on top of a common physical layer to provide on-the-fly virtualization for base stations.

\begin{figure}[!t]
\centering
    \includegraphics[scale=0.37]{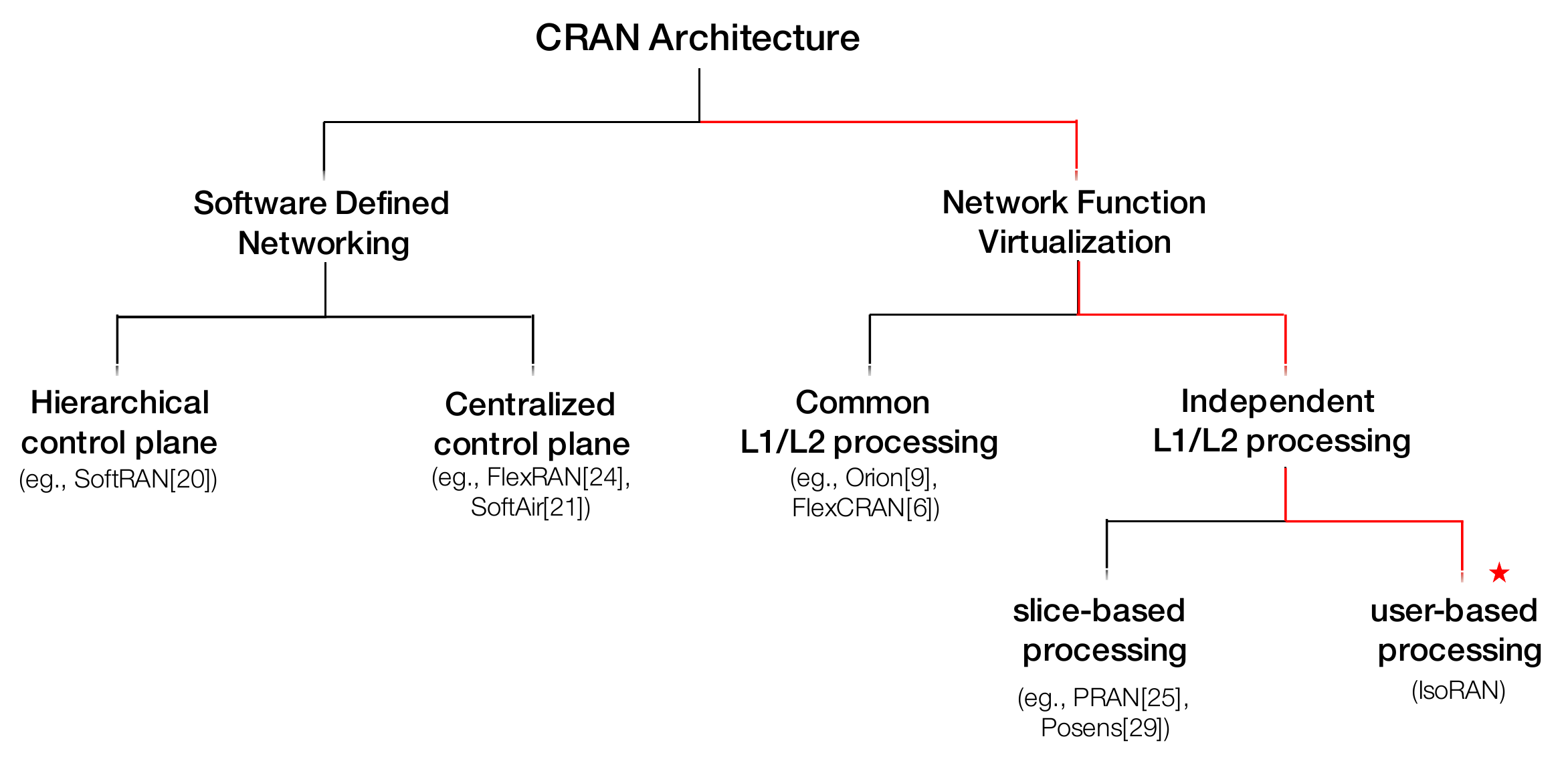} 
    \caption{Design space for CRAN architecture.}
    \vspace{-0.5cm}
    \label{fig:CranDesignSpace}
\end{figure}

A common and key limitation of the aforementioned works, which serves as one of the motivation for our work, is the lack of end-to-end slicing including the physical layer. These proposals often involve slicing upwards of L2 and hence cannot support a scenario where users are operating on frequencies which require separate data plane processing. For example, the processing for sub-6GHz range varies significantly as compared to processing for 30GHz range due the large number of antennas involved in the latter. To resolve this, multiple works~\cite{bansal2012openradio,wu2014pran,chang2018ran} introduce data plane programmability. These proposals decompose the data plane into a chain of functions where each function can be customized for a given slice. However, these works often provide limited isolation and do no scale with increasing number of slices and users per slice.

\section{Motivation}
\label{sec:Motivation}

One of the major advantages of CRAN architecture is the ability to share resources across multiple adjacent base stations. However, virtualization of baseband processing can lead to non-deterministic performance. 
Due to the tight delay constraints of one to two milliseconds, RAN processing is particularly sensitive to small variations in processing time.  
However, sharing resources such as processor, RAM, NIC cards between different slices on multiple base stations can often result in interference between different slices resulting in non-deterministic behaviors.  


Existing solutions reduce interference between base stations by placing all traffic from a single base station to a single host. In cases where multiple base stations are allocated to a single host, the allocation is done such that the host can support the worst case requirements for all the allocated base stations. To reduce inter-slice interference, most solutions propose using virtual machines for each slice to ensure isolation and reduce interference. However, static allocation of resources between virtual machines can lead to over-provisioning thereby reducing gains from multiplexing resources in the cloud. Additionally, statically allocating resources between slices is extremely difficult because of the dynamic nature of cellular networks. 

\begin{figure}[t]
\centering
    \begin{tabular}{cc}
        \includegraphics[scale=0.5]{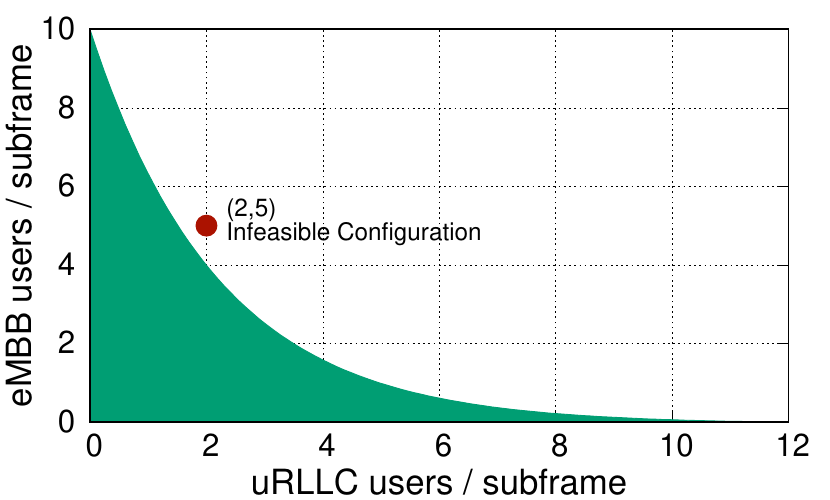} &
        \includegraphics[scale=0.5]{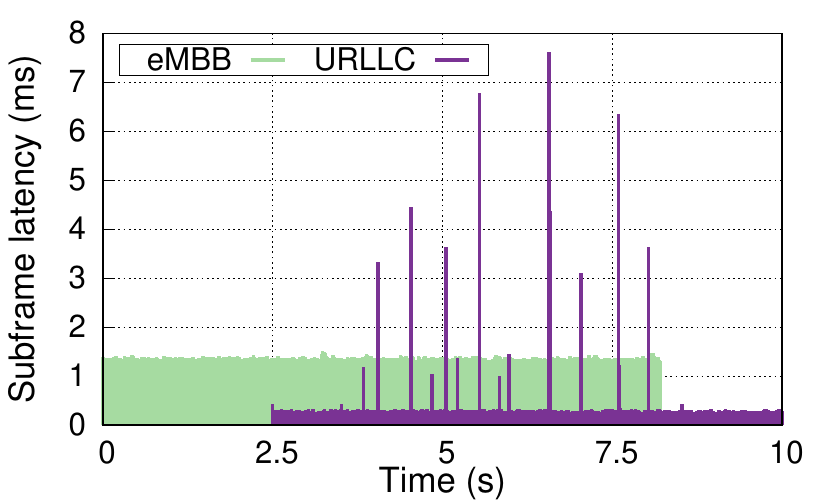}    \\        
        \multicolumn{1}{c}{\small(a) Feasible Region} & 
        \small(b) Infeasible configuration \\
    \end{tabular}
    \centering
    \caption{Interference between two slices.}
    \vspace{-0.5cm}
    \label{plot:interference}
\end{figure}



There are two 
types of interference: intra-slice and inter-slice interference. 
Intra-slice interference refers to performance degradation caused by increasing the number of users within the same slice. As the number of users are increased, the competition for shared resources such as CPU and memory on a host, results in inferior performance. Similarly, inter-slice interference is caused due to competition between slices running on the same host. 

To demonstrate the impact of each type of interference, we run a simple experiment with a CRAN setup for 5G, shown in Figure~\ref{fig:5gCranArchitecture}. Our 5G base station denoted by gNodeB supports three use cases: eMBB, uRLLC and mMTC. We simulate RAN processing for users using the PHY benchmark~\cite{phybenchmark}. 

To demonstrate intra-slice interference due to inefficient scaling we increase the number of users in a given slice. Each user has a constant traffic consumption to avoid interference arising from unpredictable network load. Figure \ref{plot:interference}(a) depicts the average data plane processing latency for a user within a slice as the number of users are increased. We see that due to the constraint on shared resources not only does the processing latency increase, the variance increases exponentially as well.

All of the proposals for RAN restrict user within a slice to the same host. Furthermore, the concept of slice is associated with each base station making infrastructure multiplexing infeasible. This leads to over-provisioning or under-provisioning at different times of the day. This is one of the guiding principles for \nameOfSystem wherein each user's allocation is independent of the base station it is attached to. Furthermore, slices are associated with the RAN overall and users from the same slice can be mapped to a host while being connected to different base stations.

We perform a simple experiment to show inter-slice interference and its impact on the performance. In this experiment we run 4 uRLLC users and 4 eMBB user on Intel-Xeon. We ensure that each slice runs on a separate set of cores with each user being assigned its own core. This ensures that the threads do not directly interfere with each other.

We run the 4 eMBB users at the beginning for 10 seconds and the uRLLC users start at 5 seconds and run for 10 seconds. However, Figure \ref{plot:interference}(b) shows that uRLLC users see a significant amount of large spikes in subframe processing latency while eMBB users are running on the same host. This implies that such critical applications can be significantly impacted even when running on a highly reliable host with sufficient processing capacity. Further proof is the complete lack of spikes once the eMBB users complete processing. This clearly shows that when two slices are running on the same host, they can interfere with each other's performance even while running on completely separate cores. 

Although recent works such as POSENS~\cite{garcia2018posens} improve isolation between slices, thereby reducing inter-slice interference, it still suffers from intra-slice interference, as it does not take into account the criticality of SLA for a slice. 
We propose a paradigm shift, wherein user processing is done separately in each thread which can be allocated anywhere in the cloud while communicating with base station thread they are attached to. 
In the next section, we discuss the main components of \nameOfSystem and its implementation.

\section{System}
\label{sec:systemdesign}



\subsection{Background}


As shown in Figure~\ref{fig:5gCranArchitecture},
Cellular Networks can be divided into three major parts: Radio Frontend\,(RF), Radio Access Network\,(RAN) and Core Network\,(CN). RF is responsible for transmitting and receiving analog signals to/from users. On the other hand, CN relays packets from the cellular network to the Internet and vice versa. It is also responsible for verifying and managing user connections. RAN sits between the RF and the CN, and is further divided into multiple layers: Physical\,(PHY), Medium Access Control\,(MAC), Radio Link Control\,(RLC), Packet Data Convergence Protocol\,(PDCP) and Radio Resource Control\,(RRC). 
These layers are responsible for various functions such as scheduling user communication, processing analog signals, interference management etc. In LTE, these layers are tightly coupled and hence are treated as a monolithic processing block.

Based on the type of function, RAN can be further divided into the control plane and the data plane. The control plane is responsible for functions such as scheduling user communication and interference management. On the other hand, the data plane is responsible for converting analog signals into bits and vice versa. In LTE, the control plane and the data plane are highly interconnected and dependent on each other. 



\begin{figure}[!t]
\centering
    \includegraphics[scale=0.24]{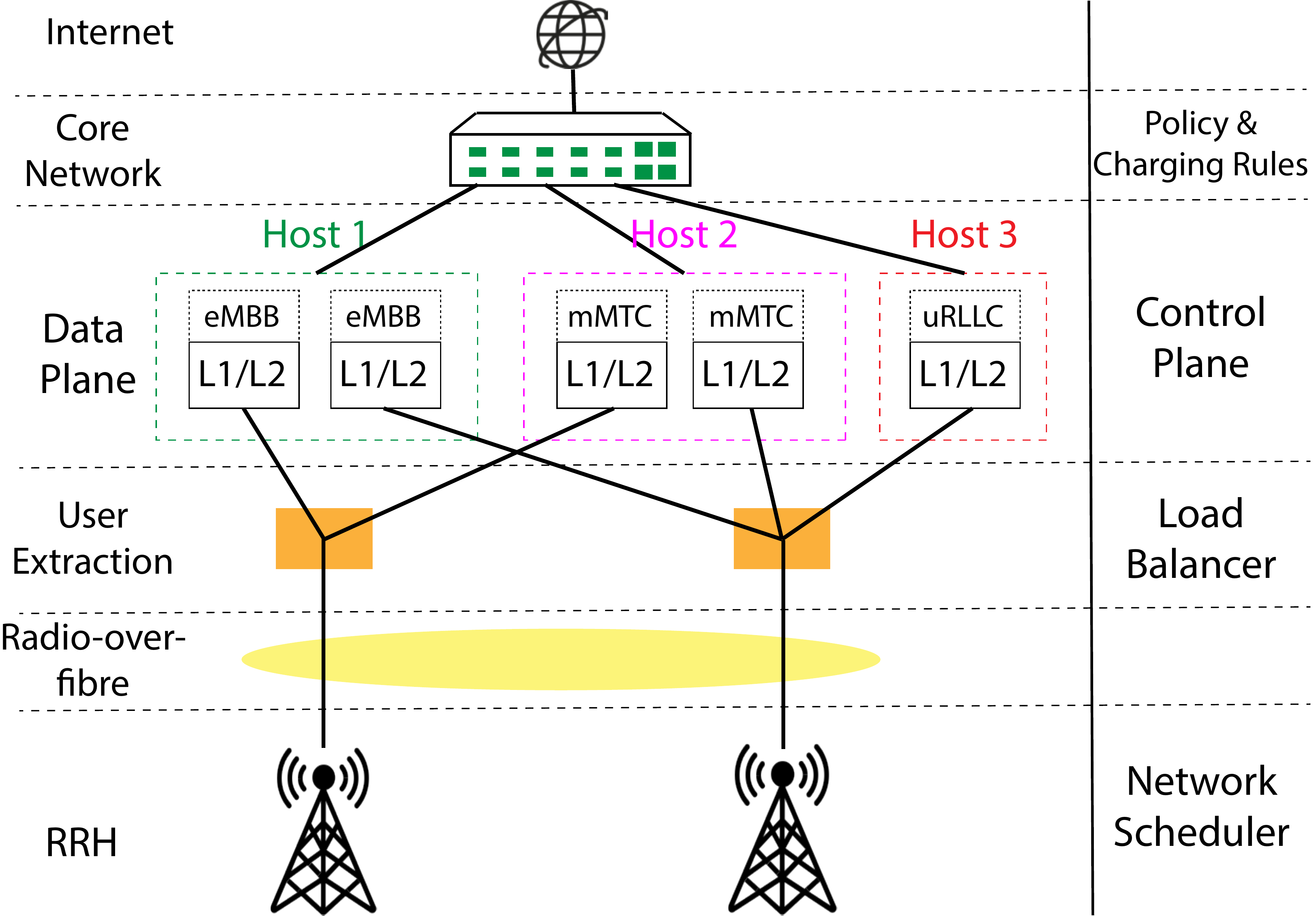}
    \caption{\nameOfSystem design overview}
    \label{fig:systemOverview}
    \vspace{-0.5cm}
\end{figure}

The data plane in the RAN consists of multiple channels in both the uplink and downlink directions. Interaction between transmissions in both directions are highly time synchronized as the ACK/NACK should be received within 4\,ms~\cite{3gpp36213} of the original transmission on the shared channel. RAN data plane processing can also be divided into user specific and cell specific functions. User specific processing involves coding and decoding data transmitted/received, which involves all the layers of LTE upto RRC. On the other hand cell specific processing deals with mostly control plane functions. 

Transmission is divided into small units along both the time and frequency domain. The smallest unit of time is called a subframe which is equal to 1\,ms. Furthermore, each subframe is divided into two slots of 0.5\,ms each. Along the frequency domain, the smallest unit is a sub-carrier which is equal to 15\,KHz. The RAN creates a new schedule every subframe. The smallest schedulable unit that can be assigned to a user is called a Physical Resource Block\,(PRB), which is a combination of twelve sub-carriers within a single slot.

\subsection{Design Overview}


We design \nameOfSystem to meet the 3 challenges mentioned in the previous section: Flexibility, Isolation and  Scalability. This section provides an overview of the main components of IsoRAN and its workings.



\begin{figure}[!t]
\centering
        \includegraphics[scale=0.2]{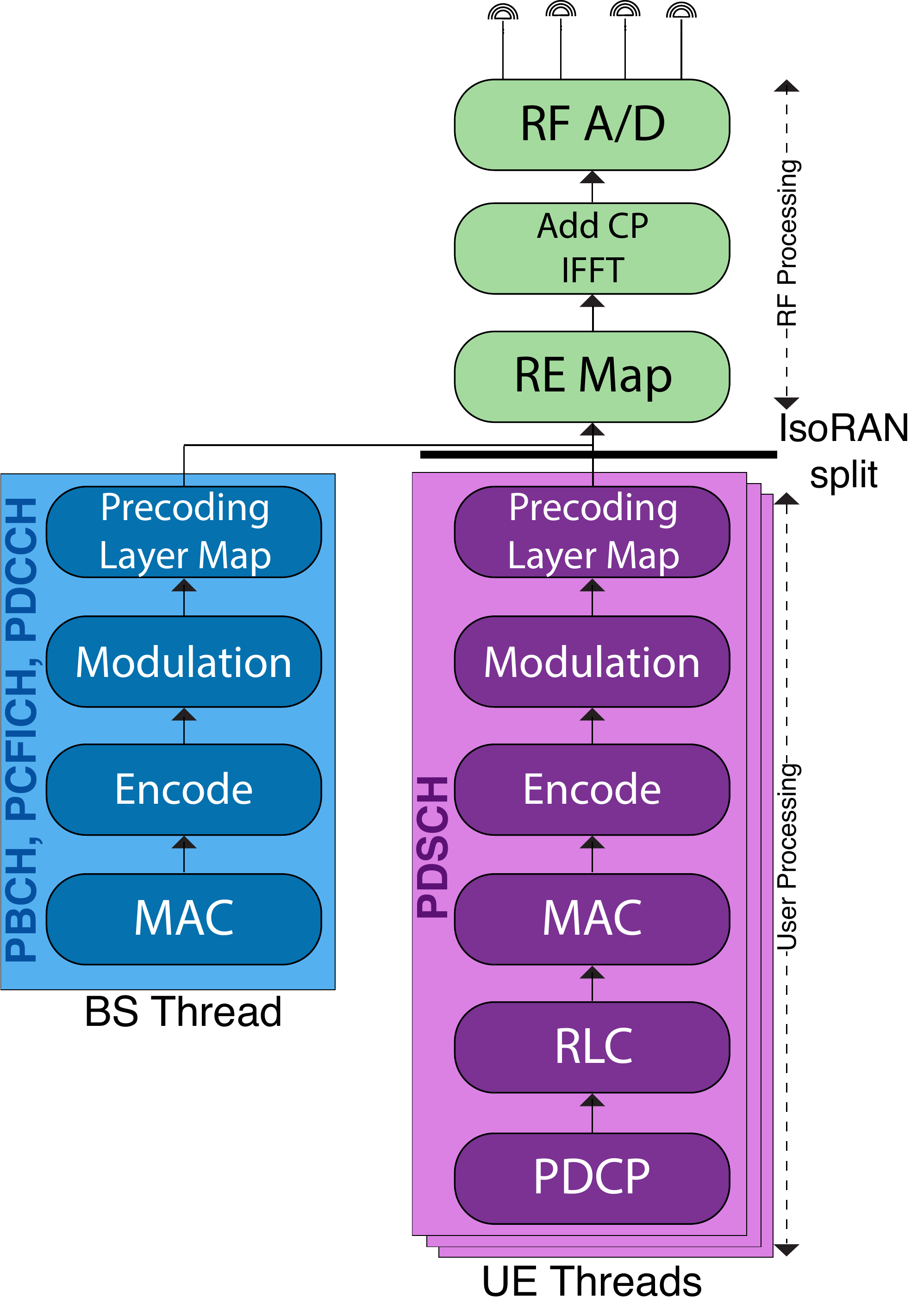} 
        \caption{User and cell specific processing for downlink}
        \vspace{-0.5cm}
        \label{fig:userCellSpecificProcessing}
\end{figure}

\textbf{Functional Split:} IsoRAN is built on top of a CRAN architecture where the IQ pairs received at the front-end are processed at a central or edge cloud. 
We choose low-PHY functional split option for \nameOfSystemNoSpace. The split option is popularly known as NGFI:IF4\cite{ngfipaper} or 3GPP:IF7-2\cite{3gpp2017study}. This split separates the common RF processing performed on the entire channel from user specific baseband processing as shown in Figure~\ref{fig:userCellSpecificProcessing} for downlink. IsoRAN can also support lower functional splits. This is important for 5G as we expect a large number of small cell installations to improve coverage.  
The common RF specific processing is performed in the base station thread before sending it to user threads for user-specific baseband processing.

\textbf{User-level Virtualization:} 
In traditional RAN architectures, the base station is the smallest unit of baseband processing. However, to support multiple use cases, the general trend has been to move towards a slice based baseband processing. Slices associated with a base station are now the smallest unit of baseband processing. IsoRAN goes further by dividing the slices down and making \emph{user-level processing as the new indivisible unit}. 

Compare to the common base station to host mapping, user-level virtualization has the following benefits. First, users with similar SLA can be placed on the same host to minimize interference. Second, user mobility becomes easier since moving from one base station to another does not necessitates moving the user's state. Finally, load balancing can be performed at a per-user level. The amount of processing needed by the CRAN can scale up and down with the total workload by simply adding new hosts rather than having to upgrade existing hosts.


We can use SIM level information to identify the user's slice to determine the custom processing required. Hence, every time a user connects to the base station, we create a thread each for its downlink and uplink processing. These threads are responsible for user specific baseband processing. Figure \ref{fig:userCellSpecificProcessing} shows the user level processing functions for downlink threads. User threads in downlink are responsible for processing the PDSCH data. The threads take IQ pairs as input and generate packets which have been processed until PDCP layer. A base station thread processes data for common channels such as PBCH, PCFICH and PUCCH which are responsible for synchronization, ACK/NACK etc. These channels are responsible for control layer communication they do not have perform higher layer processing functions for RLC and PDCP layers.


\textbf{Handling of User Mobility:} When a connection request is processed, the base station thread will update the IP address database about the new connection. For handovers, the IP addresses of the user is removed from the source base station's list and added to the target base station's list. This simplifies handovers compared to the traditional case, whereby a tunnel is setup between the base stations to handover state information. \nameOfSystem does not require such information as the state information for a user is already available in the thread. Hence, when the IP addresses are updated, the thread can continue running without requiring time consuming tasks such as setup and teardown of connecton settings (e.g. GTP tunnels). The user thread only needs to update basic static information about the base station, such as base station ID, channel bandwidth etc., to complete the handover procedure. The thread can then continue processing data which it now receives from the target base station. This reduces the time required for handover and allows for users with increased mobility to be served seamlessly without requiring any change in the UE.

Note that while user level slicing provides the above mentioned benefits, making it work requires a careful rethinking of the data and control flows in the CRAN. In particular, we need to (1) isolate user-level data plane processing such that each user thread can process its data stream independently and (2) ensures that the overhead of thread creation and communication is minimized.

\subsection{Design Details}

In order to validate our design, we have implemented a working prototype implementation of \nameOfSystem based on OAI. 
We present some of the design details and the challenges below. 

\textbf{Dependencies:}  LTE stack is highly interconnected and a considerable amount of dependencies are time constrained. In addition, with 5G, each slice will has its unique timing constraints. To solve this problem, first and foremost we separate the control and the data plane. Next, we separate the processing of each LTE channel for both uplink and downlink. We create a common thread for processing all control channels and one thread for each users PUSCH and PDSCH processing. Both these steps allow us to separate cell specific functions into either the control plane or the common base station thread. User specific processing is separated for each user and runs as its own thread. 
This design closely resembles stream processing engines which process large amounts of data in real-time. 

\begin{figure}[!t]
    \centering
    \includegraphics[scale=0.35]{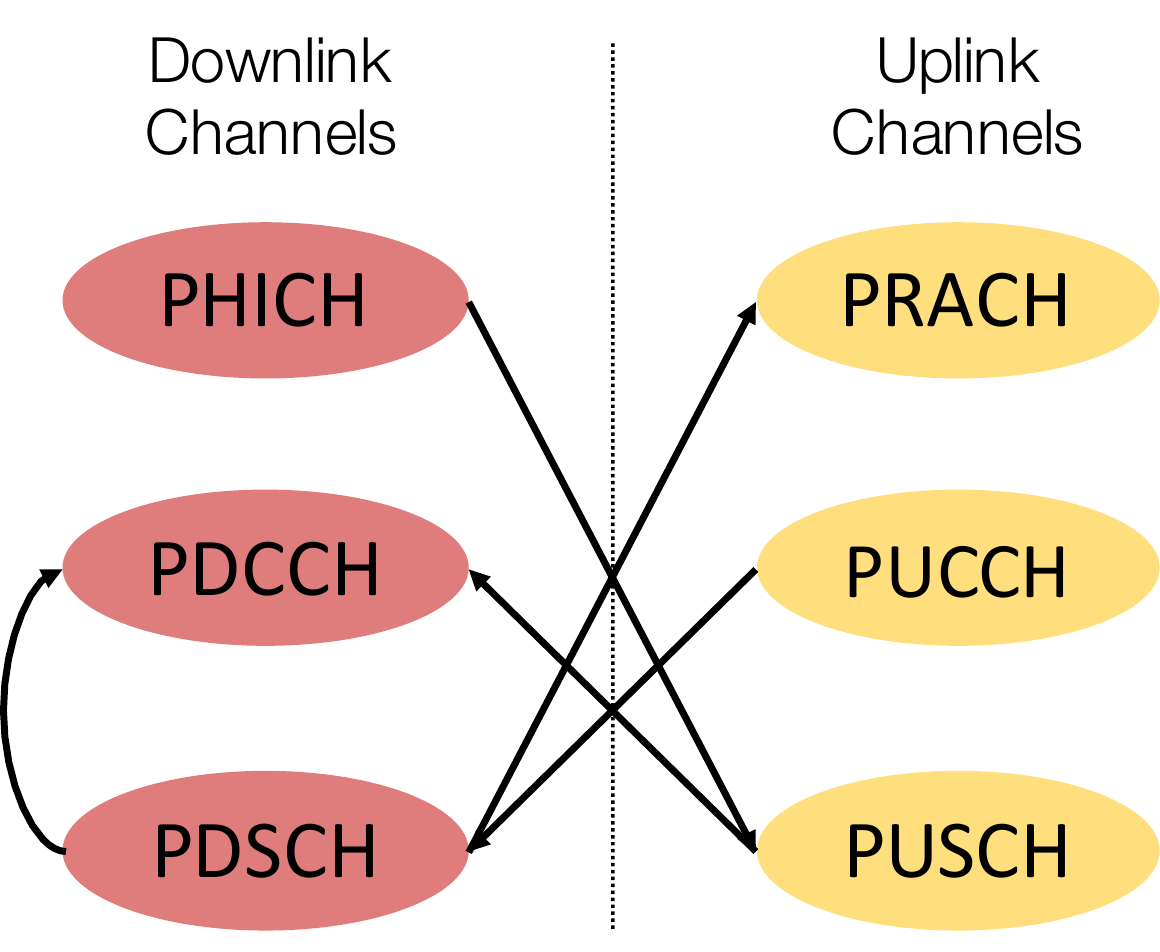}
    \caption{Dependency between LTE channels.}
    \vspace{-0.5cm}
    \label{fig:dependencyPHYChannels}
\end{figure}

\textbf{Thread communication:}
A pool of daemon is created in advanced. Hence, even though user slices are created and deleted continuously, there is no additional overhead incurred for thread creation and deletion. 
To allow inter-thread communication,
we use shared memory to store all the dependent data for the threads. This is possible because each user thread on a host has no data dependencies with other user threads. The common data shared between them such as cell ID, scheduling information is read only. The eNodeB thread is responsible for writing this shared information. Conversely, information written by the user threads are independent and do not lead to race conditions. 

IsoRAN also has to deal with the case when user threads from the same base station run on different hosts. In this case, the main eNodeB thread needs to maintain a list of IP addresses corresponding to all the UE threads attached to the base station. Furthermore, all new UE threads spawned as a result of connection request to the base station need to be added to this table. The new UE thread also needs the IP addresses of the eNodeB thread it is associated with for initial configuration and data communication thereafter. For this, we maintain a database in the centralized control plane which contains information regarding all threads. For each eNodeB thread we maintain its IP address, cell ID and list of UE threads associated with actively. For each UE thread, we maintain its IP address, associated cell ID and eNodeB thread IP address. When a new entry is added to the table the central control plane sends an update message to the eNodeB and UE thread updating their respective fields allows each to discover each other quickly.

\textbf{Splitting/Aggregating IQ Pairs: } Another part of \nameOfSystem which makes it challenging is the splitting and aggregation of IQ pair data. This is challenging because of the tight timing constraints and large size of data involved\cite{chitimalla2015reconfigurable}. For supporting a maximum cell throughput of 300\,Mbps we require a fronthaul capacity of nearly 10\,Gbps. This large size is further complicated as the splitting and aggregation changes dynamically based on schedule determined by the control plane. Furthermore, IQ pairs need to be managed in real-time to avoid affecting normal RAN processing. To manage this effectively we need to ensure that base station and user threads are allocated to threads which optimize the network. 


\textbf{UE to host allocation}
Allocating users to hosts is an important component of \nameOfSystemNoSpace, as migrating users is risky and can affect QoS at the user end.
Additionally, sub-optimal allocation not only impact the particular user but other users on the same host as well. 
For the purposes of this paper, we use a simple load balancing algorithm that uses the number of users to determine the load of each host.
However, \nameOfSystem can also support complex algorithms to produce highly optimal allocation schemes.
The discussion and evaluation of such complex algorithms is beyond the scope of this paper.

The user allocation algorithm is automatically triggered for each Random Access(RA) procedure triggered by a new user. 
Resources are allocated on a host during the transmission of Msg3 in the RA procedure when the UE requests for RRC connection. 
For each slice, we generate a priority list for host configurations available in the cloud which can support the slice requirements.
When a new user arrives, the allocation algorithm then chooses the host with the least number of users and the most optimal configuration given the user type.
For critical slices, we only allocate the user to hosts which contains the users from the same critical slice. 
If there exists no such host or there exists no resources on any viable host, we perform a similar search for the next priority configuration. 
The same search is performed for non-critical slices. 
If we do not find an appropriate host for the user, we simply request for more hosts since we are running in a cloud environment which allows dynamic resource request.




\section{Evaluation}
\label{sec:evaluation}

The evaluation is organized into three parts. 
First, we verify IsoRAN on a realistic setup using a re-engineered OAI code where multiple daemon threads are used for baseband processing as detailed in Section~\ref{sec:systemdesign}. 
We also evaluate the isolation and scaling for \nameOfSystem on the re-engineered code. 
Next, we evaluate the key principle of \nameOfSystem, per-user threading, by mapping individual user threads to processors with different capabilities. 
For this, we simulate key parts of the RAN using the PHY benchmark on three different processor architectures (2 ARM and 1 Intel). 
Finally, to evaluate the impact of \nameOfSystem on a larger scale, we execute PHY benchmark with real LTE traffic traces.
The trace has over 1.2\,million Radio Network Temporary Identifiers (RNTIs) with a total duration of 5 hours collected over 15 days.

\begin{figure}[!t]
\centering
    \includegraphics[width=7.2cm,height=4.75cm]{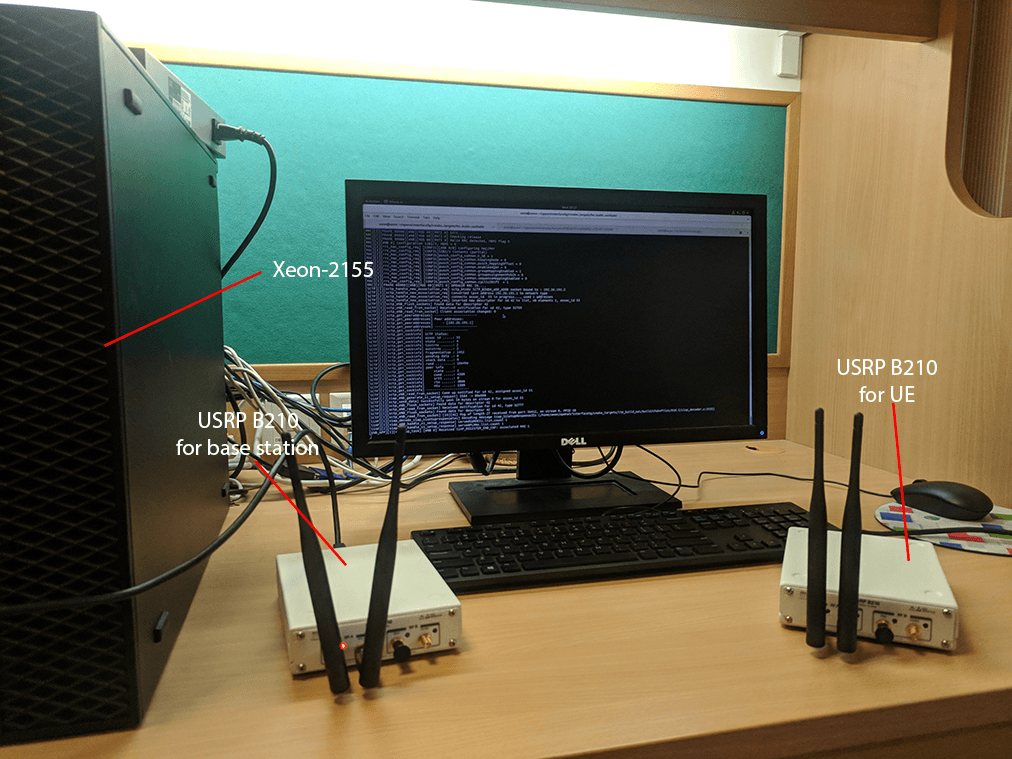} 
    \caption{Setup of OAI with SDR and base stations.}
    \label{fig:oaisetup}
    \vspace{-0.5cm}
\end{figure}

\begin{figure}[!t]
\centering
    \begin{tabular}{cc}
        \includegraphics[scale=1]{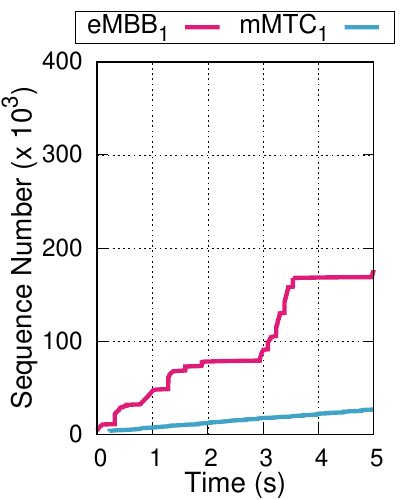}      &
        \includegraphics[scale=1]{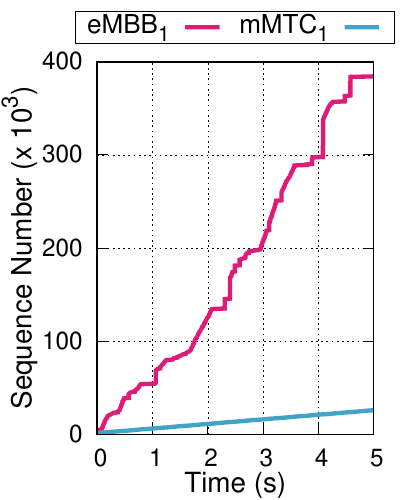}  \\
        (a) Vanilla OAI & (b) \nameOfSystem  \\
    \end{tabular}
    \centering
    \caption{Comparison of number of Bytes transmitted by users of different slices for Vanilla OAI and \nameOfSystem}
    \vspace{-0.5cm}
    \label{plot:oaiMultipleSlices}
\end{figure}



\subsection{Comparison to Vanilla OpenAirInterface (OAI)}


\textbf{Working Prototype:} We use the setup as shown in Figure~\ref{fig:oaisetup} to verify \nameOfSystem in an end-to-end setup.
We use a USRP B210 for the RF and OpenAir-CN for the core network.
We use a commodity server with has a Xeon-2155 processor having 10 cores and a maximum frequency of 3.3\,GHz for baseband processing.
The server runs \nameOfSystem with each core running a daemon thread to process user data transmitted/received in each subframe.
We disable hyperthreading and other power management tools for all our experiments for better accuracy. 
To verify the system's functionality, we establish a two way communication, using ping packets, between the mobile user and public servers such as Google. 

To compare the memory and processing overhead in \nameOfSystem we use OpenAirInterface's in-built emulator.
The emulator provides synthetic radio channels while still executing the entire LTE stack.
We recorded the memory footprint and CPU utilization of the server when running Vanilla OAI and IsoRAN when multiple users are sending data to public servers. 
To ensure similar user workload, each user sends UDP packets with a constant bitrate of 400Kbps.

Table \ref{table:compareIsoranWithVanillaOai} shows that \nameOfSystem consumes slightly higher memory compared to Vanilla OAI because of the extra memory needed to communicate between different threads. 
However, we see that the overhead remains constant, as the overhead is independent of number of users in the system and depends only on RAN properties such as channel bandwidth, sampling frequency, bits per sample etc. 
Additionally, average core utilization is relatively low for \nameOfSystemNoSpace, as we spread the processing across multiple cores. 
This is important as \nameOfSystem scales with increasing number of cores as compared to contemporary works which are limited to a single core or a single host. 
This is crucial as Vanilla OAI experienced severe degradation in QoS when 3 UEs are active, often leading to packet drops caused by its inability to meet timing constraints imposed by the LTE stack.
For a single user, \nameOfSystem has slightly higher core utilization due to the overhead caused by thread communication. 
However, this overhead is negligible as compared to the total processing as the number of users in the RAN increases.
The result shows that IsoRAN is a viable solution with minimal overhead.

\begin{table}[!t] 
\centering
\setlength{\tabcolsep}{5pt}
    \caption[]{Comparing Memory Footprint and CPU Utilization for IsoRAN over a traditional designs. The CPU Utilization is averaged over active cores.}
    \begin{tabular}{|p{1cm}|c|c|c|c|}
        \hline
                    & UEs   & Memory (GB)   & Core & Avg. Core Util.   \\ \hline \hline
        \multirow{4}{*}{\parbox{1cm}{\centering Vanilla \\ OAI}}
                &    No UE  &  1.11         & 1 & 35.14\%           \\ \cline{2-5}
                &    1      &  1.68         & 1 & 48.67\%           \\ \cline{2-5}
                &    2      &  2.25         & 1 & 52.16\%           \\ \cline{2-5}
                &    3*      &  2.82         & 1 & 59.00\%           \\ \hline \hline
        \multirow{4}{*}{IsoRAN}
                &    No UE  &  1.19         & 1 & 35.67\%           \\ \cline{2-5}
                &    1      &  1.76         & 2 & 24.67\%           \\ \cline{2-5}
                &    2      &  2.33         & 3 & 18.47\%           \\ \cline{2-5}
                &    3      &  2.90         & 4 & 21.82\%           \\ \hline
    \end{tabular}
    \label{table:compareIsoranWithVanillaOai}
    \vspace{-0.75cm}
\end{table}

\textbf{Isolation:}
To evaluate isolation, we consider two users, one eMBB and one IoT (or mMTC) user. 
To simulate eMBB users we use iperf connections, with TCP protocol, to saturate the wireless bandwidth. 
The IoT user sends a 400 byte ping packet every 100\,ms to simulate transmissions which are small and bursty in nature. 
Note that the additional processing load imposed by the ping (IoT) traffic compared to the iperf (eMBB) traffic is very small.

From the figure, for Vanilla OAI, it is clear that the progress of eMBB user is significantly affected due to the presence of the IoT user  which is bursty in nature and does not have throughput management schemes. 
Again, such adverse interference occurs even though the increase in CPU utilization is very small. For the IoT user, the maximum RTT exceeds 350\,ms due to resource sharing with the eMBB user. 
On the other hand, With IsoRAN, by assigning these services to different cores, the performance of both users are not affected by each other.

In the following sections, we will show that in realistic scenarios with more users, \nameOfSystem can achieve good performance in terms of isolation with less resources.

\vspace{-0.2cm}

\subsection{Scaling on Heterogeneous Platforms}

A key benefit of \nameOfSystem is that with user-level threading, we can place users from different slices on different hosts that better match its processing requirements. 
To show the effectiveness of \nameOfSystem, we test it for 3 slices on 3 different host configurations. 
Due to the issue with scaling and portability with OAI, we do not perform the evaluation using OAI, but instead evaluate using PHY benchmark\cite{phybenchmark} which simulates ULSCH processing. 
We modify PHY benchmark to ensure the presence of daemon threads for baseband processing and each thread is affined to a core. 
For each active user within a subframe, we assign it to a unique free daemon thread for its baseband processing.
We also instrument the code to use a custom trace and return subframe processing latency.

Each user within a slice is assumed to have the maximum possible capacity defined by 3GPP specifications and available on state-of-the-art hardware.
Therefore each eMBB user uses 64QAM modulation scheme and 4x4 transmissions. 
One of possible use cases under mMTC is IoT communication.
Therefore, to estimate the network usage we use a combination theoretical values from the 3GPP specifications and existing state-of-the-art IoT device capabilities~\cite{extensisIot}\cite{3gpp36213}. 
In our case study, the maximum speed for IoT devices is set to 50\,Kbps with QPSK modulation. 
Finally, for uRLLC users we use the Autonomous Vehicle Tracking\,(AVT) as the application, where each user sends small updates continuously with the total throughput of over 2\,Mbps.

\begin{figure}[!t]
\centering
    \begin{tabular}{cc}
        \includegraphics[width=3.25cm,height=4cm]{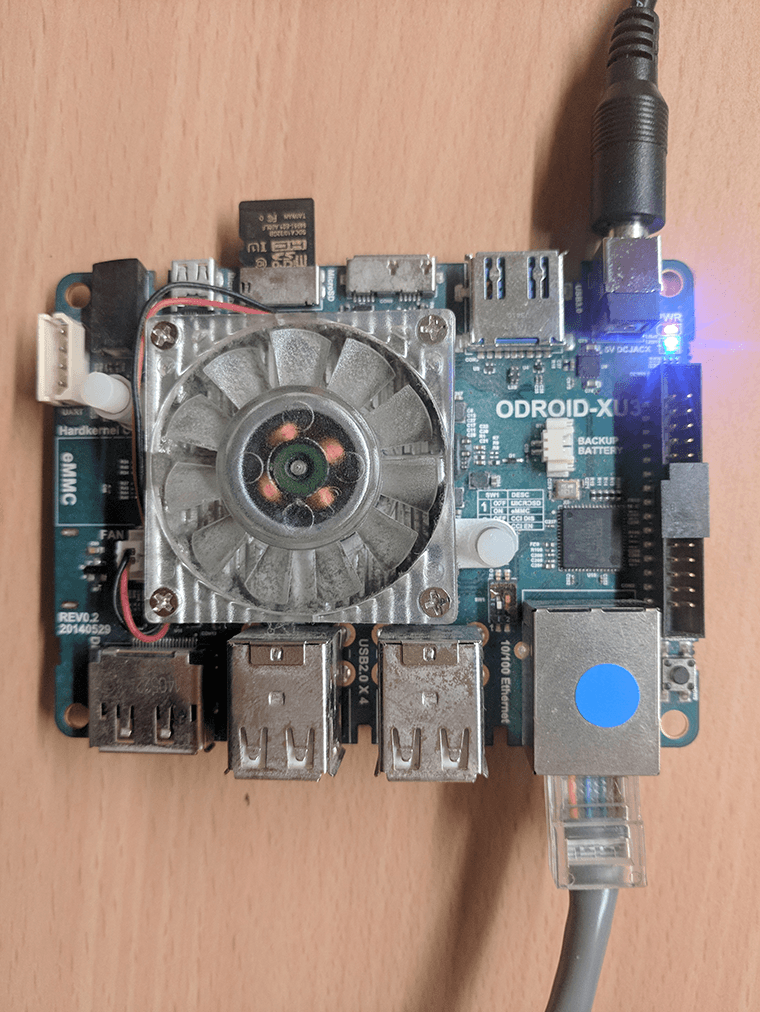}      &
        \includegraphics[width=3.25cm,height=4cm]{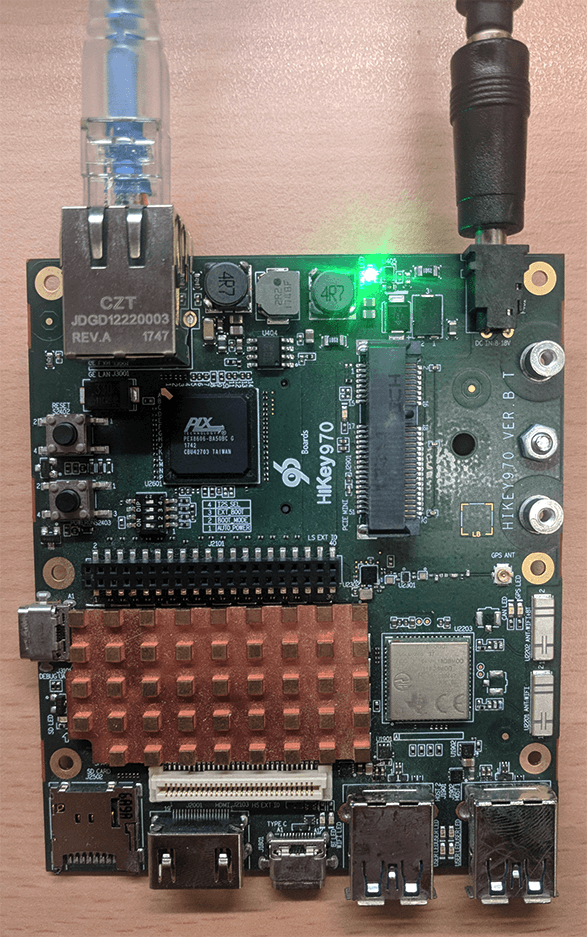}  \\
        (a) OdroidXU3 & (b) HiKey 970  \\
    \end{tabular}
    \centering
    \caption{Boards used for A7 and A73 respectively}
    \vspace{-0.5cm}
    \label{figure:boards}
\end{figure}

\begin{table}[!t] 
\centering
\setlength{\tabcolsep}{5pt}
    \caption[]{Specs of different processors used in evaluation}
    \begin{tabular}{|c|c|c|c|}
        \hline
                        & Odroid XU3    & HiKey 970     & Xeon-2155     \\ 
                        & (A7)          & (A73)         &               \\ \hline \hline
        Freq. (MHz)     & 1400          & 2340          & 3300          \\ \hline     
        Cores           & 4             & 4             & 10            \\ \hline     
        Price (USD)     & 60            & 320           & 1440          \\ \hline  
    \end{tabular}
    \vspace{-0.5cm}
    \label{table:processorDetails}
\end{table}

We use 3 different types of processors: ARM A7, ARM A73 and Xeon-2155 (see Table~\ref{table:processorDetails}).
The Xeon emulate the typical high power processors commonly used in data centers. The 
ARM processors emulate processing capability available in modern-day network interface cards (NICs) and can be used to support mMTC and uRLLC users\cite{netronomeNIC,mellanoxNIC,caviumNIC}.
There are multiple works that used these processors on smartNICs for executing microservices to improve overall performance and achieve lower latency~\cite{liu2019e3}.
ARM A7 is a power efficient in-order core whereas ARM A73 is a more powerful out-of-order execution architecture. 
We use the OdroidXU3 board which contains the A7 and the HiKey board which contains A73 (see figure \ref{figure:boards}), to obtain power and CPU utilization values. 
To measure power consumption on Intel Xeon-2155, we use Running Average Power Limit\,(RAPL) which provides energy status counters that track energy consumption.

\subsubsection{Worst Case Execution}

\begin{table}[!t] 
\centering
\setlength{\tabcolsep}{3pt}
    \caption[]{Worst Case Overhead, Latency, CPU Utilization and Power consumed by each use case on different configuration}
    \begin{tabular}{|c|c|c|c|c|}
        \hline
         Use Case & Comm. Overhead & ARM A7 & ARM A73 & Xeon-2155  \\ \hline \hline
        \multirow{3}{*}{mMTC} & \multirow{3}{*}{11us}
         & 0.81ms        & 0.30ms        & 0.28ms        \\
         & & 62.36\%      & 27.34\%       & 2.4\%          \\ 
         & & 0.59W        & 3.32W         & 82.98W    \\ \hline \hline
        \multirow{3}{*}{uRLLC} & \multirow{3}{*}{15us}
         & NA due to    & 0.48ms        & 0.39ms    \\
         & & latency      & 37.75\%       & 3.2\%          \\ 
         & & constraints  & 3.79W         & 85.22W      \\ \hline \hline
        \multirow{3}{*}{eMBB} & \multirow{3}{*}{407us}
         & NA due to    & NA due to        & 2.35ms    \\
         & & latency      & latency       & 86.9\%          \\ 
         & & constraints  & constraints   & 102.67W      \\ \hline
    \end{tabular}
    \vspace{-0.5cm}
    \label{table:compareSlicesOnMultipleArchitecture}
\end{table}

For completeness, we show the worst case execution behavior of running three different service types on three different processor in Table \ref{table:compareSlicesOnMultipleArchitecture}. 
Clearly, the lower power ARM A7 can only support low bit rate service such as mMTC, while the more powerful ARM A73 can support mMTC and uRLLC but not eMBB. Xeon-2155 can support all service types. 
Using this information, we can construct a priority list for each slice by ordering the feasible processors based on their power consumption and feasibility.
Therefore, the priority list for mMTC is A7, A73 and Xeon-2155; uRLLC is A73 and Xeon-2155; and eMBB is only Xeon-2155.

The overhead also includes the transmission of IQ pairs from the base station thread to the user thread. 
We measure the worst case by transmitting the maximum size of data for each slice over a network with multiple hops connected by 10Gbps and that the processor is placed 4 hops away from the base station. 
From Table \ref{table:compareSlicesOnMultipleArchitecture} we can see that the overhead is in the range of microseconds for both mMTC and uRLLC. 
For eMBB even though the overhead is high it is still much smaller as compared the required latency bound. 
We expect the network overhead to become negligible with the introduction of multi-point routing in the recent release of eCPRI\cite{cpri2019ecpri}.

\subsubsection{User Thread Allocation}

\begin{table}[!t] 
\caption{Slice allocation on different hosts for schemes.}
\centering
\setlength{\tabcolsep}{5pt}
    \begin{tabular}{|c|c|c|c|}
        \hline
        Scheme  & ARM A7    & ARM A73           & Xeon-2155     \\ \hline \hline
        I       & 2 mMTC    & 2 uRLLC           & 5 eMBB          \\ \hline     
        II      & -         & 2 mMTC, 2 uRLLC   & 5 eMBB          \\ \hline     
        III     & -         & 2 uRLLC           & 2 mMTC, 5 eMBB          \\ \hline  
        IV      & 2 mMTC    & -                 & 2 uRLLC, 5 eMBB          \\ \hline  
        V  & \multirow{2}{*}{-}& \multirow{2}{*}{-}    & 2 mMTC, 2 uRLLC          \\ 
        (Baseline)    &           &                   & 5 eMBB          \\ \hline  
    \end{tabular}
    \vspace{-0.5cm}
    \label{table:userallocationforschemes}
\end{table}

\begin{figure}[!t]
\centering
    \includegraphics[scale=0.9]{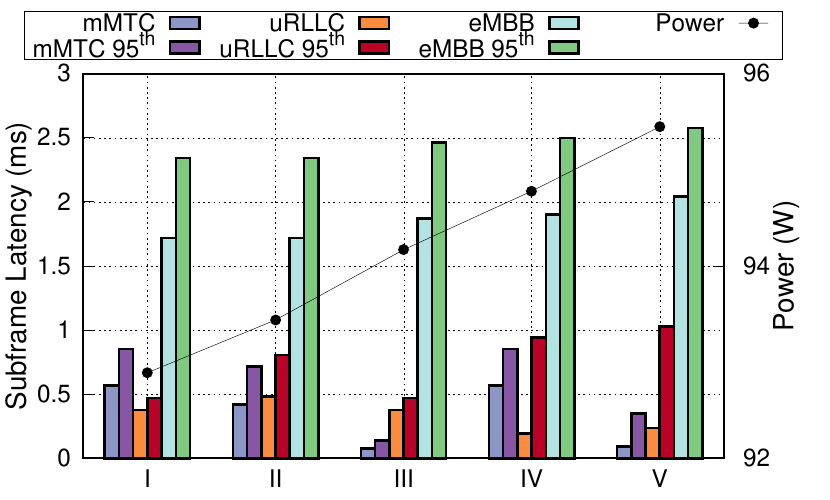}
    \vspace{-0.5cm}
    \caption{Subframe latency for each slice plotted against total power consumption. Allocation for different schemes described in Table \ref{table:userallocationforschemes}.}
    \label{plot:latencyfordifferentcases}
    \vspace{-0.5cm}
\end{figure}

\begin{table*}[!htbp]
\centering
\setlength{\tabcolsep}{5pt}
    \caption{Comparison of costs for different scenarios for large-scale simulation.}
     \begin{tabularx}{0.96\linewidth}{@{\extracolsep{\fill}}|r|Y|Y|Y|Y|Y|Y|Y|Y|Y|}
    \hline
         & \multicolumn{3}{c|}{Base Case} & \multicolumn{3}{c|}{\nameOfSystem Homogeneous} & \multicolumn{3}{c|}{\nameOfSystem Heterogeneous} \\ \hline
    No. of base stations & 3 & 15 & 30 & 3 & 15 & 30 & 3 & 15 & 30 \\ \hline \hline
    Xeon Cores & 30 & 150 & 300 & 30 & 100 & 160 & 20 & 90 & 140 \\ \hline
    ARM A73 Cores & NA & NA & NA & 0 & 0 & 0 & 4 & 12 & 24\\ \hline
    ARM A7 Cores & NA & NA & NA & 0 & 0 & 0 & 4 & 12 & 24 \\ \hline 
    Cost\,(\$) & 4320 & 21600 & 43200 & 4320 & 16200 & 24480 & 3225 & 13900 & 20800 \\ \hline
    Power\,(W) & 256.5 & 1280 & 2564.2 & 256 & 960.85 & 1453.58 & 174.24 & 760.68 & 1132.96 \\ \hline
    mMTC Drop Rate(\%) & 
    \multicolumn{3}{c|}{3.39} &  \multicolumn{3}{c|}{0.002} & \multicolumn{3}{c|}{0.002}   \\ \hline
    eMBB Drop Rate(\%) & \multicolumn{3}{c|}{4.59} &  \multicolumn{3}{c|}{3.32} &  \multicolumn{3}{c|}{3.36}  \\ \hline
    uRLLC Drop Rate(\%) & \multicolumn{3}{c|}{42.7} &  \multicolumn{3}{c|}{0.005} & \multicolumn{3}{c|}{4.58}   \\ \hline
    \end{tabularx}
    \vspace{-0.5cm}
    \label{table:comparisonLargeScale}
\end{table*}

In order to evaluate the gain when using heterogeneous processors, we used 5 different thread allocation schemes to support the load of 2 mMTC, 2 uRLLC and 5 eMBB users running at maximum workload. 
IsoRAN's allocation is shown as allocation scheme I, where users from each slice are placed on the most optimal host that matches the processing requirement. 
In the other three schemes (II- IV), we remove one of the hosts from scheme 1. 
For example, in scheme II, we do not use the A7 host and put both the mMTC and uRLLC on the same ARM A73.
In scheme V, we show the baseline case of using only 1 Xeon-2155 host.
Table \ref{table:userallocationforschemes} shows the 5 allocation schemes used and the results are summarized in Figure \ref{plot:latencyfordifferentcases}. 

Given that, scheme I uses the most resources, one would expect scheme 1 to have the highest power consumption. However, our results show that this is not the case and in fact the reverse is true.
By putting the slices on separate hosts, IsoRAN has the lowest power consumption as well as the lowest latency values, both for average and the $95^{th}$ percentile in many cases.
In scheme II, when the A7 host is removed and both the uRLLC and mMTC users run on the A73 host, due to inter-slice interference between uRLLC and mMTC, the latency sensitive uRLLC user has a much higher $95^{th}$ percentile while consuming more power. Scheme III puts mMTC and eMBB on Xeon-2155 which results in mMTC users have significantly lower processing latency. However, this lower subframe latency comes at the cost of higher power and higher average and $95^{th}$ latency for eMBB users. Scheme IV puts both uRLLC and eMBB users on the Xeon-2155. While the average subframe latency for uRLLC users drops significantly, the $95^{th}$ percentile latency values increases as compared to previous schemes implying that many more subframes exceed the 0.5ms latency threshold. This echoes the results from Figure \ref{plot:interference} where placing critical and non-critical users on the same host degrades performance for critical users.

Finally in Scheme V we run users of all slices on the Xeon-2155 host. We see that due to inter-slice interference the average latency of eMBB and the 95$^{th}$ percentile latency of uRLLC slices go up. Furthermore, because Xeon-2155 is a power hungry core, executing all users on it leads to the highest power consumption of any possible allocation scheme.


In summary, we have shown that the following. First, the inability to separate slices on separate hosts causes performance degradation due to interference. This interference adversely impacts critical time-sensitive slices significantly. Second, the availability of heterogeneous processors allow more flexibility to place the processing on optimal hardware, thus reducing overall cost and power while meeting the service requirements. 

\subsection{Large Scale Simulation}

To show the benefits of IsoRAN with respect to scaling, we perform a large-scale simulation using real traces captures from network operators. To capture these traces we use IMDEA's Online Watcher for LTE~\cite{bui2016owl}. We capture 10 minute traces from adjacent base stations over an extended period of 15 days. Each trace captures the user's scheduling information such as resource block allocation, modulation scheme and coding rate for each subframe. We use this information to simulate the load on the base station using PHY benchmark.

In the simulation, for the purpose of slicing, we classify users into 3 different slices based on their traffic patterns. 
uRLLC slice users are identified as users with medium data rate constant with at most 
8\,RBs per frame. 
Users with Transport Block Size (TBS) less than 1000\,bits for all subframes are classified as mMTC based on the Transport Block Size (TBS) table for NB-IoT communication provided in 3GPP specs~\cite{3gpp36213}.
The rest of the users are classified as eMBB users.

We run a baseline where all the users of a base station are run on a single host. 
For IsoRAN we separate users from different slice and allocate hosts using the simple allocation algorithm. 
To evaluate the benefit of user level virtualization alone we perform a set of experiments using only Xeon-2155 hosts while ensuring isolation between difference slice.
Additionally, to evaluates the benefits of user-level virtualization combined with the usage of hosts with different capabilities, we perform another set of experiments using heterogeneous processor architectures.
The Heterogeneous case uses a combination of ARM A7, ARM A73 and Xeon-2155 for processing users from different slices.
For both the cases in \nameOfSystem we the allocation algorithm as described in Section~\ref{sec:systemdesign}.
Our baseline executes all users connected to a base station on a Xeon-2155 host as existing RAN architectures do not facilitate spilling users from the same base station to different hosts. 

Table \ref{table:comparisonLargeScale} shows the drop rates for different slices for the three cases.
We calculate the percent of subframes whose processing exceeds the subframe latency threshold.
For eMBB, the drop rate reduces by 25\% on average for both homogeneous and heterogeneous IsoRAN. 
The reduction in \nameOfSystem comes due to the absence of users from other slices (uRLLC and mMTC) which do no compete for resource with eMBB users.
The main advantage comes in reduction of drop rate for uRLLC users from over 40\% to nearly 0\% for the homogeneous case. 
The heterogeneous case uses ARM Cortex-A73 and the drop rate is 4.58\%. 
Note that, the drop rate for uRLLC users on A73 is higher than recommended rate, because the processor is underpowered for our classification in this simulation. 
In practice, one can easily address this by using a more powerful ARM processors such as the ARM Cortex-A76, which will reduce the drop rate significantly with minimal increase in cost and power.
Finally, mMTC users, see significantly lower latency for both the homogeneous and heterogeneous cases.

Table \ref{table:comparisonLargeScale} illustrates savings for user-level virtualization for both operating costs, i.e. power consumption and fixed costs, i.e. equipment cost. 
We see that as the number of base station increases both the homogeneous and heterogeneous cases, the savings increases because of user-level virtualization. 
Power savings increases from 32\% to 55\% for Heterogeneous case over the baseline as we move from 3 base stations to 30. 
Similarly, the cost savings also increases from 25\% to 52\%.



\section{Conclusion}
\label{sec:conclusion}

In this paper we show how existing slicing proposals do not address isolation and scaling sufficiently. We show that dividing the network into slices is not sufficient as it leads to inter and intra slice interference. This adversely affects performance for critical slices. To avoid such interference we propose IsoRAN that virtualizes user plane RAN processing. This enables per-user orchestration and improves user allocation to hosts. 
We have built a working prototype of IsoRAN on OpenAirInterface.
We experimentally shows that IsoRAN insulates users within the same slice from interfering with each other. We also show that the overhead for distributing RAN processing on heterogeneous set of hosts is negligible. The overhead can be further minimized for time-sensitive slices by smart placement of hosts in the cloud.

%

\end{document}